# Discrete ellipsoidal statistical BGK model and Burnett equations

Yudong Zhang[1,2], Aiguo Xu[2,3*], Guangcai Zhang[2], Zhihua Chen[1]

Pei Wang[2]

1 Key Laboratory of Transient Physics, Nanjing University of Science and Technology, Nanjing 210094, China

2 Laboratory of Computational Physics, Institute of Applied Physics and Computational Mathematics, Beijing, 100088, China

3 Center for Applied Physics and Technology, MOE Key Center for High Energy Density Physics Simulations, College of Engineering, Peking University, Beijing 100871, China

*Corresponding author. E-mail: Xu_Aiguo@iapcm.ac.cn

## Abstract

A new discrete Boltzmann model, discrete Ellipsoidal Statistical(ES)-BGK model, is proposed to simulate non-equilibrium compressible flows. Compared with the original discrete BGK model, the discrete ES-BGK has a flexible Prandtl number. For the discrete ES-BGK model in Burnett level, two kinds of discrete velocity model are introduced; the relations between non-equilibrium quantities and the viscous stress and heat flux in Burnett level are established. The model is verified via four benchmark tests. In addition, a new idea is introduced to recover the actual distribution function through the macroscopic quantities and their space derivatives. The recovery scheme works not only for discrete Boltzmann simulation but also for hydrodynamic ones, for example, based on the Navier-Stokes, the Burnett equations, etc.

**Key words:** discrete Boltzmann model, ellipsoidal statistical BGK, Burnett equations, non-equilibrium quantities, actual distribution function.

## 1. Introduction

Rarefied gas flows are traditionally associated with spacecraft re-entry into planetary atmosphere where the air is so thin that the applicability of Navier-Stokes (NS) model is challenged[1-3]. Recently, the rarefied effect of flows in microchannels have attracted significant research interest due to the rapid development of micro-fluidic technologies such as Micro-Electro-Mechanical System(MEMS)[4,5]. Generally, the rarefaction of flows can be measured by a dimensionless parameter, the Knudsen number (Kn), which is defined as the ratio of the mean free path of molecules to characteristic length that we focus on. In summary, there are two types of rarefied gas flows, one is the thin gas which has a large molecular distance, such as the air in the high altitude atmosphere[1], and the other is the flows with small characteristic length, such as shock wave and MEMS. In fact, according to the value of Kn, the flow can be divided into four categories including continuum flow(Kn<0.001), slip flow (0.001<Kn<0.1), transitional flow(0.1<Kn<10), and free molecular flow (Kn>10)[1,4,5].

As we know, NS equations are applicable to continuum and slip flow (with slip boundary conditions), but it fails to provide the correct viscous stress and heat flux in transitional regime. The reason for the inapplicability of NS equations in transitional flow is that the constitutive equations, i.e. Newton's viscosity law and the Fourier heat conduction law, are assumed to be linear which is inapposite when the non-equilibrium (or rarefaction) effect is significant. The Burnett equations[3,6], which are obtained from the Boltzmann equation through Chapman-Enskog (CE) expansion, have a modified constitutive equations and can work in part of the transition flow zones. However, the Burnett equations often encounter numerical instabilities because of the high order derivatives in the viscosity and heat flux terms[6].

It has been known that Boltzmann equation is applicable for all of the four flow regimes mentioned above. Unfortunately, the original Boltzmann equation is too complicated to be solved directly[7]. The multidimensional nature of distribution function and collision operator pose a great challenge for its numerical solution. Realistic numerical computations of the Boltzmann equation are based on probabilistic methods, such as direct simulation Monte Carlo (DSMC) method[8], or deterministic fast numerical methods, such as fast spectral method (FSM)[7]. In general, however, the computation cost is still too expensive for the direct solution of the Boltzmann equation. So, a variety of simplified methods have been developed to approximate the solution of the Boltzmann equation, such as the unified gas-kinetic scheme(UGKS)[9,10], the discrete velocity method (DVM)[11], the discrete unified gas-kinetic scheme (DUGKS)[12,13], the lattice Boltzmann method (LBM)[14-18], and the discrete Boltzmann method(DBM)[19-21].

Recently, DBM has been developed as a non-equilibrium flow simulation tool and has been widely used in various flow conditions including high speed compressible flow, multiphase flow[22], flow instability[23-24], combustion and detonation[25-26], etc. Besides the values and evolutions of conserved kinetic moments, the DBM presents also those of the most relevant non-conserved kinetic moments. The latter is helpful for understanding the constitutive relations for the former. Some of the new observations brought by DBM, for example, the non-equilibrium fine structures of shock waves [27], have been confirmed and supplemented by the results of molecular dynamics[28-30]. It should be pointed out that the molecular dynamics simulations can also give microscopic view of points to the origin of the slip near boundary, such as the non-isotropic strong molecular evaporation flux from the liquid[31], which might help to develop more physically reasonable mesoscopic models for slip-flow regime.

Generally, there are two steps to simplify the full Boltzmann equation. The first step is to simplify the collision operator. Several kinds of simplified collision operators including BGK model[32], ellipsoidal statistical BGK (ES-BGK) model[33], Shakhov model[34], Rykov model[35], etc. are presented to substitute the collisional integral of the Boltzmann equation. Among those models, the BGK model is most extensively used because of its simplicity. But the BGK model fails to give the proper Prandtl number (Pr) since the Pr in BGK model is fixed to unity. The ES-BGK model, presented by Holway[33], employing a Gaussian distribution as the relaxation equilibrium state instead of the Maxwell distribution, is a modification of the BGK and possesses an adjustable Prandtl number. More importantly, the validity of the H-theorem for the ES-BGK model has been proved by Andries and his collaborators[36]. In addition, Zheng and Struchtrup[37] have found that the Burnett equations derived from the ES-BGK model completely agree with the Burnett equations from the full Boltzmann equation for Maxwell molecules. So, the ES-BGK model is more preferable when the Pr effect is concerned.

The second step to simplify the Boltzmann equation is to discrete the particle velocity space. According the modeling idea of discrete Boltzmann method (DBM), the velocity space can be substituted by a limited number of particle velocities on condition that the specific kinetic moments remain unchanged[19, 20]. The choice of those kinetic moments depends on the specific physical problem and technically via the CE analysis. For example, only the $0^{th}$ to $4^{th}$ orders of velocity kinetic moments are needed for the DBM in NS level. Inspired by the previous work about DBM[38], we consider to present a framework of discrete ES-BGK model, which contains the multiscale characteristic of DBM and can give a proper Prandtl number. As an example, the discrete ES-BGK model in Burnett level is demonstrated and verified by numerical simulation.

The remainder of this paper is organized as follows. Section 2 demonstrates the derivation process of the ES-BGK model from the original Boltzmann equation. Section 3 shows the modeling of discrete ES-BGK including evolution equation, the required kinetic moments, and the discrete velocity model. Section 4 presents the relation between new model and the Burnett equation, and the relations between non-equilibrium quantities and the constitutive equations in Burnett equations. Section 5 gives the simulation results of the new model for four test cases. Section 6 provides a new method to recovery the actual velocity distribution function from the DBM approximately but quantitatively. Section 7 concludes the present paper.

## 2. Ellipsoidal statistical BGK model

The original Boltzmann equation of single-component gases reads[2]

$$\frac{\partial f}{\partial t} + \mathbf{v} \cdot \frac{\partial f}{\partial \mathbf{r}} + \mathbf{a} \cdot \frac{\partial f}{\partial \mathbf{v}} = \left(\frac{\partial f}{\partial t}\right)_c, \quad (1)$$

where $\mathbf{a}$ is the external force per unit mass, and $\left(\frac{\partial f}{\partial t}\right)_c$ is the collisional integral and can be represented as

$$\left(\frac{\partial f}{\partial t}\right)_c = \int_{-\infty}^{\infty} \int_0^{4\pi} \left(f'(\mathbf{v}_*)f'(\mathbf{v}) - f(\mathbf{v}_*)f(\mathbf{v})\right) c_r \sigma d\Omega d\mathbf{v}_*, \quad (2)$$

where $f(\mathbf{v}_*)$ and $f(\mathbf{v})$ indicate the velocity distribution functions of two pre-collision molecules, and $f'(\mathbf{v}_*)$ and $f'(\mathbf{v})$ are the corresponding velocity distribution functions of the two post-collision molecules, $c_r = |\mathbf{v}_* - \mathbf{v}|$ is the relative pre-collision velocity and $\sigma$ is the collision cross section.

To simplify the collisional integral, the $K(f)$ is introduced[33]

$$K(f) = \int_{-\infty}^{\infty} \int_0^{4\pi} f(\mathbf{v}_*) c_r \sigma d\Omega d\mathbf{v}_* . \tag{3}$$

Then, the collisional integral in Eq.(2) can be represented as

$$\left(\frac{\partial f}{\partial t}\right)_c = -K(f)[f(\mathbf{v}) - \psi(f)], \tag{4}$$

where $\psi(f)$ is functional of $f$ and will be determined later. In fact, the physical meaning of the term $-K(f)f(\mathbf{v})d\mathbf{v}d\mathbf{r}$ is the number of "absorption" molecules by collisions from the range $d\mathbf{v}d\mathbf{r}$ around the point $(\mathbf{v},\mathbf{r})$ in phase space per unit time, while $K(f)\psi(f)d\mathbf{v}d\mathbf{r}$ is that of the "emission" molecules by collisions. The collision term needs to satisfy the conservation of mass, momentum, and energy, so it has

$$\int_{-\infty}^{\infty} \left(\frac{\partial f}{\partial t}\right)_c \theta(\mathbf{v}) d\mathbf{v} = 0, \tag{5}$$

where $\theta(\mathbf{v}) = 1$, $\mathbf{v}$, or $\mathbf{v} \cdot \mathbf{v}$. If we assume $K(f)$ to be independent of molecular velocity[33], the relations in Eq.(5) for three dimensions are equivalent to

$$\int_{-\infty}^{\infty} \psi(\mathbf{v}) d\mathbf{v} = \rho, \tag{6}$$

$$\int_{-\infty}^{\infty} \psi(\mathbf{v}) \mathbf{v} d\mathbf{v} = \rho \mathbf{u}, \tag{7}$$

$$\int_{-\infty}^{\infty} \psi(\mathbf{v})(\mathbf{v} - \mathbf{u}) \cdot (\mathbf{v} - \mathbf{u}) d\mathbf{v} = 3\rho RT. \tag{8}$$

In order to determine $\psi(\mathbf{v})$ being subjected to requirements of Eqs.(6)-(8), the entropy $S$ is introduced which can be used to measure the amount of uncertainty[33]. The $S$ is defined as

$$S = -\int \psi(\mathbf{v}) \ln \psi(\mathbf{v}) d\mathbf{v}. \tag{9}$$

If no further information about $\psi(\mathbf{v})$ is known beyond conditions Eqs.(6)-(8), the most possible assignment of the $\psi(\mathbf{v})$ is the one which maximizes the uncertainty $S$ subject to these conditions. Using the Lagrangian multiplier method, one can obtain the form of $\psi(\mathbf{v})$ as Maxwellian distribution,

$$\psi(\mathbf{v}) = \rho \frac{1}{(2\pi RT)^{3/2}} \exp\left[-\frac{|\mathbf{v}-\mathbf{u}|^2}{2RT}\right], \tag{10}$$

when $S$ gets its maximum. Go a further step, let us consider that more restrictions are added to $\psi(\mathbf{v})$, namely,

$$\int_{-\infty}^{\infty} \psi(\mathbf{v})(v_\alpha - u_\alpha)(v_\beta - u_\beta) d\mathbf{v} = \rho \lambda_{\alpha\beta}, \tag{11}$$

where $\lambda_{\alpha\beta}$ is assumed to be known and the specific form of $\lambda_{\alpha\beta}$ will be given later.

Under the conditions of Eq.(6), Eq.(7), and Eq.(11), when S gets its maximum, one can obtain the form of $\psi(\mathbf{v})$ as

$$\psi(\mathbf{v}) = \rho \frac{1}{(2\pi)^{3/2} |\lambda_{\alpha\beta}|^{1/2}} \exp\left[-\frac{1}{2}\lambda_{\alpha\beta}^{-1}(v_\alpha - u_\alpha)(v_\beta - u_\beta)\right], \tag{12}$$

which is the so-called ellipsoidal statistical model because $\psi(\mathbf{v})$ in Eq.(12) is an ellipsoidal distribution in velocity space.

Now, the specific form of $\lambda_{\alpha\beta}$ are to be determined. Firstly, the conservation of energy requires Eq.(8) to be satisfied so that the trace of $\lambda_{\alpha\beta}$ is

$$\lambda_{\alpha\alpha} = 3RT = \frac{1}{\rho} M^*_{\alpha\alpha} \tag{13}$$

For convenient, $\lambda_{\alpha\beta}$ is chosen as a linear function of the second-order moments of $f$,

$$\lambda_{\alpha\beta} = \frac{1}{\rho} G_{\alpha\beta\gamma\lambda} M^*_{\gamma\lambda} \tag{14}$$

where $M^*_{\alpha\beta} = \int_{-\infty}^{\infty} f(\mathbf{v})(v_\alpha - u_\alpha)(v_\beta - u_\beta) d\mathbf{v}$ is the second-order central moment. Since the collision integral is an isotropic operator, $G_{\alpha\beta\gamma\lambda}$ must be an isotropic tensor. The

most general form for the isotropic tensor of four order can be represented as

$$G_{\alpha\beta\gamma\lambda} = a_1\delta_{\alpha\beta}\delta_{\gamma\lambda} + a_2\delta_{\alpha\gamma}\delta_{\beta\lambda} + a_3\delta_{\alpha\lambda}\delta_{\beta\gamma}. \tag{15}$$

Since $M^*_{\alpha\beta}$ is symmetric, Eq.(14) becomes

$$\lambda_{\alpha\beta} = \frac{1}{\rho}\left(a_1 M^*_{\gamma\gamma}\delta_{\alpha\beta} + bM^*_{\alpha\beta}\right), \tag{16}$$

where $b = a_2 + a_3$. From the relation of Eq.(13), it has $a_1 = \frac{1-b}{3}$. So one can obtain

$$\lambda_{\alpha\beta} = (1-b)RT\delta_{\alpha\beta} + \frac{b}{\rho}M^*_{\alpha\beta} = RT\delta_{\alpha\beta} + \frac{b}{\rho}\tilde{M}^*_{\alpha\beta}, \tag{17}$$

where $\tilde{M}^*_{\alpha\beta} = M^*_{\alpha\beta} - \frac{1}{3}M_{\gamma\gamma}\delta_{\alpha\beta}$ is the traceless symmetric tensor of $M^*_{\alpha\beta}$.

From Eqs.(1), (4), (12), and (17) the ellipsoidal statistical model is obtained. However, there are still two parameters to be determined including $K$ in Eq.(4) and $b$ in Eq.(17). From the CE multiscale expansion[33,39] and the comparison with NS equations, one can obtain

$$b = \frac{\Pr-1}{\Pr}, \tag{18}$$

and

$$K = \Pr\frac{\rho RT}{\mu}. \tag{19}$$

where Pr is the Prandtl number and $\mu$ is the viscosity coefficient. In the latter part of this article, for convenient, the $K$ is taken as $\frac{1}{\tau}$ where $\tau$ is the reciprocal of the molecular collision frequency.

## 3. Discrete ellipsoidal statistical BGK model

From the previous derivation, we know the evolution equation of ES-BGK model reads

$$\frac{\partial f}{\partial t} + v_\alpha \frac{\partial f}{\partial r_\alpha} = -\frac{1}{\tau}(f - f^{ES}), \tag{20}$$

when external forces are omitted, where $f^{ES}$ is an ellipsoidal distribution as given in

Eq.(12) with $\lambda_{\alpha\beta}=RT\delta_{\alpha\beta}+\dfrac{b}{\rho}\tilde{M}^*_{\alpha\beta}$. When $b=0$ (i.e. Pr=1), $f^{ES}$ recovers to Maxwellian distribution $f^{eq}$ which has a form as shown Eq.(10). In that case, the ellipsoidal statistical model recovers to the BGK model.

Combining the Non-Organized Momentum Flux defined in Ref.[26], one can find that $\tilde{M}^*_{\alpha\beta}$ is essentially equivalent to $\Delta^*_{2,\alpha\beta}$ which corresponds to the viscous stress tensor in the macroscopic constitutive equations. So the expression of $\lambda_{\alpha\beta}$ can be rewritten as

$$\lambda_{\alpha\beta}=RT\delta_{\alpha\beta}+\dfrac{b}{\rho}\Delta^*_{2,\alpha\beta}. \tag{21}$$

The velocity space is substituted by a limited number of particle velocities, and the distribution function $f$ is replaced by $f_i$ where the subscript indicates the index of the discrete velocities, then the evolution equation of discrete ES-BGK model is obtained

$$\dfrac{\partial f_i}{\partial t}+v_\alpha\dfrac{\partial f_i}{\partial r_{i\alpha}}=-\dfrac{1}{\tau}(f_i-f_i^{ES}). \tag{22}$$

From the CE expansion, hydrodynamic equations, including Euler, NS, and BGK-Burnett equations, can be obtained from Eq.(20). Specifically, the CE expansion indicates that, in order to obtain NS and BGK-Burnett equations, the $0^{th}$ to $4^{th}$ orders and the $0^{th}$ to $5^{th}$ orders of the velocity kinetic moments of $f^{eq}$ and $f^{ES}$ are needed, respectively. In other words, only if those several orders of the velocity kinetic moments of $f^{eq}$ and $f^{ES}$ in integral form can be equally calculated from the corresponding summation form of $f_i^{eq}$ and $f_i^{ES}$ about particle velocities, can the NS (BGK-Burnett) equations be derived from Eq.(22). The $0^{th}$ to $5^{th}$ orders of the velocity kinetic moments of $f_i^{ES}$, $M_m^{ES}$ $(m=0,\cdots,5)$, for BGK-Burnett equations, are listed below

$$M_0^{ES}=\sum_{i=1}^{N}f_i^{ES}=\rho \tag{23}$$

$$M_{1,\alpha}^{ES}=\sum_{i=1}^{N}f_i^{ES}v_{i\alpha}=\rho u_\alpha \tag{24}$$

$$M^{ES}_{2,\alpha\beta} = \sum_{i=1}^{N} f_i^{ES} v_{i\alpha} v_{i\beta} = \rho\left(J_{\alpha\beta} + u_\alpha u_\beta\right) \tag{25}$$

$$M^{ES}_{3,\alpha\beta\gamma} = \sum_{i=1}^{N} f_i^{ES} v_{i\alpha} v_{i\beta} v_{i\gamma} = \rho\left(u_\alpha J_{\beta\gamma} + u_\beta J_{\alpha\gamma} + u_\gamma J_{\beta\alpha} + u_\alpha u_\beta u_\gamma\right) \tag{26}$$

$$M^{ES}_{4,\alpha\beta\gamma\chi} = \sum_{i=1}^{N} f_i^{ES} v_{i\alpha} v_{i\beta} v_{i\gamma} v_{i\chi} = \rho(J_{\alpha\beta\gamma\chi} + J_{\alpha\beta} u_\gamma u_\chi + J_{\alpha\gamma} u_\beta u_\chi + J_{\alpha\chi} u_\beta u_\gamma + J_{\beta\gamma} u_\alpha u_\chi + J_{\beta\chi} u_\alpha u_\gamma + J_{\gamma\chi} u_\alpha u_\beta + u_\alpha u_\beta u_\gamma u_\chi) \tag{27}$$

$$M^{ES}_{5,3\alpha\beta\gamma} = \sum_{i=1}^{N} f_i^{ES} v_{i\alpha} v_{i\beta} v_{i\gamma} \frac{v_{i\chi}^2}{2} = \frac{1}{2} M^{ES}_{5,\alpha\beta\gamma\chi\chi} \tag{28}$$

where $J_{\alpha\beta} = \lambda_{\alpha\beta}$, $J_{\alpha\beta\gamma\xi} = \lambda_{\alpha\beta}\lambda_{\gamma\xi} + \lambda_{\alpha\gamma}\lambda_{\beta\xi} + \lambda_{\alpha\xi}\lambda_{\beta\gamma}$, and $\lambda_{\alpha\beta}$ can be found in Eq.(21). Besides, the third order tensor $M^{ES}_{5,\alpha\beta\gamma\chi\chi}$ is contracted from the $M^{ES}_{5,\alpha\beta\gamma\lambda\xi}$ which reads

$$\begin{aligned}
M^{ES}_{5,\alpha\beta\gamma\lambda\xi} = \rho(&J_{\alpha\beta\gamma\lambda} u_\xi + J_{\alpha\beta\gamma\xi} u_\lambda + J_{\alpha\beta\lambda\xi} u_\gamma + J_{\alpha\gamma\lambda\xi} u_\beta + J_{\beta\gamma\lambda\xi} u_\alpha \\
&+ J_{\alpha\beta} u_\gamma u_\lambda u_\xi + J_{\alpha\gamma} u_\beta u_\lambda u_\xi + J_{\alpha\lambda} u_\beta u_\gamma u_\xi + J_{\alpha\xi} u_\beta u_\gamma u_\lambda + J_{\beta\gamma} u_\alpha u_\lambda u_\xi \\
&+ J_{\beta\lambda} u_\alpha u_\gamma u_\xi + J_{\beta\xi} u_\alpha u_\gamma u_\lambda + J_{\gamma\lambda} u_\alpha u_\beta u_\xi + J_{\gamma\xi} u_\alpha u_\beta u_\lambda + J_{\lambda\xi} u_\alpha u_\beta u_\gamma \\
&+ u_\alpha u_\beta u_\gamma u_\lambda u_\xi)
\end{aligned}$$

The kinetic moments of $f_i^{eq}$, $M_m$ $(m=0,\cdots,5)$, can be also obtained from Eqs.(23)-(28) when $b=0$ in $\lambda_{\alpha\beta}$. Actually, those kinetic moment equations can be written in a matrix form, i.e.,

$$\mathbf{C} \cdot \mathbf{f}^{ES} = \mathbf{M}^{ES} \tag{29}$$

where $\mathbf{C}$ is the coefficient matrix whose elements are determined by discrete velocities, $\mathbf{f}^{ES}$ is a vector of discrete velocities distribution function whose number of components equals to the number of discrete velocities, and $\mathbf{M}^{ES}$ is a vector of the kinetic moments.

It can be seen, once the discrete velocities are determined, the form of the coefficient matrix $\mathbf{C}$ is known. The matrix $\mathbf{C}$ is a square matrix when the number of discrete velocities exactly equals to the number of independent kinetic moments, otherwise it is not a square matrix. Nevertheless, once the discrete velocities are known, the discrete ellipsoidal statistical distribution function $f_i^{ES}$ can be solved from Eq.(29). The choice of the discrete velocity depends on the compromise among the following several points: (i) numerical efficiency, (ii) numerical stability, and (iii) to which extent the local symmetry should be kept. The third point relies on the specific physical problem under consideration.

In this study, as an example, two kinds of discrete velocity model (DVM) for two-dimensional (2D) case are presented. Considering that the number of independent kinetic moments is 19 in Eqs.(23)-(28) for two-dimensional (2D) case, the first DVM contains 19 discrete velocities and is denoted by D2V19. From the above discussion we know the coefficient matrix **C** is square for this kind DVM. In this case, the square matrix **C** must be full rank to ensure that Eq. (29) has a solution. So the discrete velocities model is chosen as Fig.1(a) and the special value of each discrete velocity is given in Eq.(30). The other DVM contains 36 discrete velocities and is denoted by D2V36 which is shown in Fig.1(b). and the special value of corresponding discrete velocity can be found in Eq.(31).

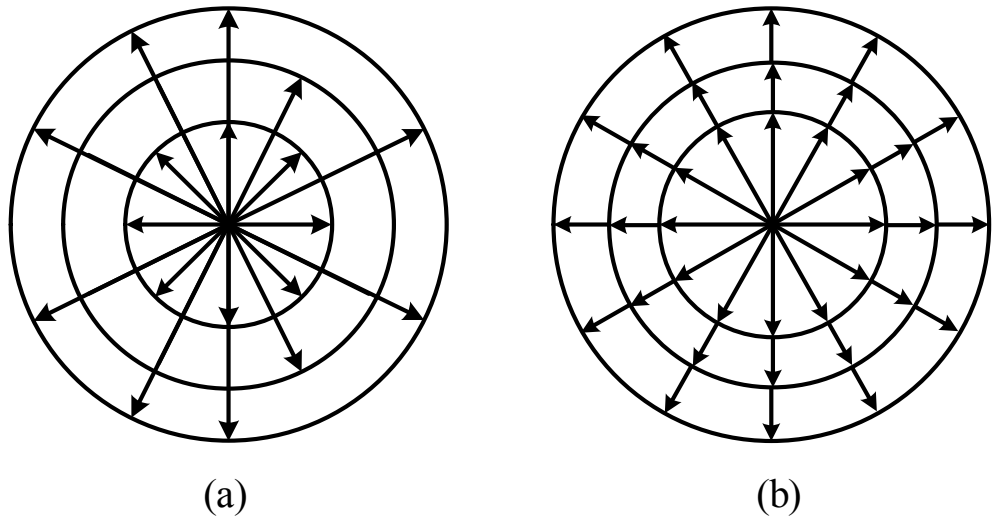

(a)          (b)

Fig.1 Schematic of discrete velocity model. (a) D2V19 model, (b) D2V36 model.

$$\mathbf{v}=(v_{ix},v_{iy})=\begin{cases}(0,0), & i=1\\ c\left(\cos\dfrac{(i-2)\pi}{4},\sin\dfrac{(i-2)\pi}{4}\right), & i=2-9\\ a\left(\cos\dfrac{(i-9)\pi}{6},\sin\dfrac{(i-9)\pi}{6}\right) & \begin{cases}a=2c, & i=10-14\ \&\ i\neq 11\\ a=\dfrac{3}{2}c, & i=11\end{cases}\\ a\left(\cos\dfrac{(i-8)\pi}{6},\sin\dfrac{(i-8)\pi}{6}\right) & \begin{cases}a=2c, & i=15-19\ \&\ i\neq 18\\ a=\dfrac{3}{2}c, & i=18\end{cases}\end{cases} \quad (30)$$

$$\mathbf{v} = (v_{ix}, v_{iy}) = \begin{cases} c\left(\cos\dfrac{(i-1)\pi}{6}, \sin\dfrac{(i-1)\pi}{6}\right), & i = 1-12 \\ 2c\left(\cos\dfrac{(i-13)\pi}{6}, \sin\dfrac{(i-13)\pi}{6}\right), & i = 13-24 \\ 3c\left(\cos\dfrac{(i-25)\pi}{6}, \sin\dfrac{(i-25)\pi}{6}\right), & i = 25-36 \end{cases} \quad (31)$$

where $c$ is an adjustable parameter.

## 4. Ellipsoidal statistical BGK model and Burnett equations

Since the velocity distribution function can be expanded around its local thermodynamic equilibrium state

$$f_i = f_i^{eq} + \varepsilon f_i^{(1)} + \varepsilon^2 f_i^{(2)} + \cdots, \quad (32)$$

and the $f^{ES}$ can also be expanded around the local equilibrium distribution function

$$f_i^{ES} = f_i^{eq} + \varepsilon f_i^{ES(1)} + \varepsilon^2 f_i^{ES(2)} + \cdots, \quad (33)$$

the general equations of hydrodynamics can be obtained from Eq.(22), which read[26]

$$\frac{\partial \rho}{\partial t} + \frac{\partial (\rho u_\alpha)}{\partial r_\alpha} = 0, \quad (34)$$

$$\frac{\partial (\rho u_\alpha)}{\partial t} + \frac{\partial (\rho u_\alpha u_\beta + p\delta_{\alpha\beta})}{\partial r_\beta} + \frac{\partial}{\partial r_\beta} \Delta^*_{2,\alpha\beta} = 0, \quad (35)$$

$$\frac{\partial (\rho E)}{\partial t} + \frac{\partial [(\rho E + p)u_\alpha]}{\partial r_\alpha} + \frac{\partial}{\partial r_\alpha}\left(u_\beta \Delta^*_{2,\alpha\beta} + \Delta^*_{3.1,\alpha}\right) = 0, \quad (36)$$

where the non-equilibrium quantities $\Delta^*_{2,\alpha\beta}$ and $\Delta^*_{3.1,\alpha}$ are the so called Non-organized momentum flux (NOMF) and Non-organized energy flux (NOEF).

At the NS level, $\Delta^*_{2,\alpha\beta}$ and $\Delta^*_{3.1,\alpha}$ are approximately equal to $\Delta^{*(1)}_{2,\alpha\beta}$ and $\Delta^{*(1)}_{3.1,\alpha}$, respectively. That is

$$\Delta^*_{2,\alpha\beta} \approx \Delta^{*(1)}_{2,\alpha\beta} = \sum_{i=1}^{N} \varepsilon f_i^{(1)} (v_{i\alpha} - u_\alpha)(v_{i\beta} - u_\beta), \quad (37)$$

$$\Delta^*_{3.1,\alpha} \approx \Delta^{*(1)}_{3.1,\alpha} = \frac{1}{2}\sum_{i=1}^{N} \varepsilon f_i^{(1)} (v_{i\gamma} - u_\gamma)^2 (v_{i\alpha} - u_\alpha), \quad (38)$$

where $N$ is the number of discrete velocities.

From the CE expansion, the $f_i^{(1)}$ can be repressed by $f_i^{eq}$,

$$\varepsilon f_i^{(1)} = -\tau \left( \varepsilon \frac{\partial f_i^{eq}}{\partial t_1} + v_\alpha \varepsilon \frac{\partial f_i^{eq}}{\partial r_{1\alpha}} \right) + \varepsilon f_i^{ES(1)}, \tag{39}$$

where time multi-scale expansion has been adopted, i.e. $\frac{\partial}{\partial t} = \varepsilon \frac{\partial}{\partial t_1} + \varepsilon^2 \frac{\partial}{\partial t_2} + \cdots$, $\frac{\partial}{\partial r_\alpha} = \varepsilon \frac{\partial}{\partial r_{1\alpha}}$, $t_1$ and $t_2$ are two independent time scales. Then we can obtain

$$\Delta_{2,\alpha\beta}^* \approx \Delta_{2,\alpha\beta}^{*(1)} = 2\mu \frac{\partial u_{<\alpha}}{\partial r_{\beta>}}, \tag{40}$$

$$\Delta_{3.1,\alpha}^* \approx \Delta_{3.1,\alpha}^{*(1)} = \kappa \frac{\partial T}{\partial r_\alpha}, \tag{41}$$

where $\mu = \frac{1}{1-b} \tau p$, $\kappa = c_p \tau p$, $c_p$ is the specific heat ratio at constant pressure, and $\frac{\partial u_{<\alpha}}{\partial r_{\beta>}}$ is the trace-free tensors which reads $\frac{\partial u_{<\alpha}}{\partial r_{\beta>}} = \frac{1}{2}(\frac{\partial u_\alpha}{\partial r_\beta} + \frac{\partial u_\beta}{\partial r_\alpha}) - \frac{1}{D} \frac{\partial u_\gamma}{\partial r_\gamma} \delta_{\alpha\beta}$, where $D$ is the spatial dimension[39].

In the Burnett level, the $\Delta_{2,\alpha\beta}^*$ and $\Delta_{3.1,\alpha}^*$ are approximately equal to $\Delta_{2,\alpha\beta}^{*(1)} + \Delta_{2,\alpha\beta}^{*(2)}$ and $\Delta_{3.1,\alpha}^{*(1)} + \Delta_{3.1,\alpha}^{*(2)}$, respectively,

$$\Delta_{2,\alpha\beta}^* \approx \Delta_{2,\alpha\beta}^{*(1)} + \Delta_{2,\alpha\beta}^{*(2)}, \tag{42}$$

$$\Delta_{3.1,\alpha}^* \approx \Delta_{3.1,\alpha}^{*(1)} + \Delta_{3.1,\alpha}^{*(2)}, \tag{43}$$

where $\Delta_{2,\alpha\beta}^{*(2)}$ and $\Delta_{3.1,\alpha}^{*(2)}$ read

$$\Delta_{2,\alpha\beta}^{*(2)} = \sum_{i=1}^N \varepsilon^2 f_i^{(2)} \left( v_{i\alpha} - u_\alpha \right)\left( v_{i\beta} - u_\beta \right), \tag{44}$$

$$\Delta_{3.1,\alpha}^{*(2)} = \frac{1}{2} \sum_{i=1}^N \varepsilon^2 f_i^{(2)} \left( v_{i\gamma} - u_\gamma \right)^2 \left( v_{i\alpha} - u_\alpha \right). \tag{45}$$

Similarly, $f_i^{(2)}$ can be repressed by $f_i^{(1)}$,

$$\varepsilon^2 f_i^{(2)} = -\tau \left( \varepsilon^2 \frac{\partial f_i^{eq}}{\partial t_2} + \varepsilon \frac{\partial f_i^{(1)}}{\partial t_1} + v_\alpha \varepsilon \frac{\partial f_i^{(1)}}{\partial r_\alpha} \right) + \varepsilon^2 f_i^{ES(2)}, \tag{46}$$

where $f_i^{(1)}$ can be further substituted by $f_i^{eq}$ using Eq.(39).

If the viscosity $\mu$ is adopted as a function of temperature[37], the relaxation time $\tau$ has a relation with macroscopic quantities which reads

$$\tau = (1-b)\frac{\mu}{p} = \tau_0 \rho^{-1} T^{\beta-1}, \quad (47)$$

where $\tau_0 = \frac{(1-b)\mu_0}{T_0^\beta}$ is the reference relaxation time, the $\Delta_{2,\alpha\beta}^{*(2)}$ and $\Delta_{3.1,\alpha}^{*(2)}$ can be obtained as the same form as the Eq.(28) in literature[37]. As a preliminary study, however, we will take the relaxation time $\tau$ as a constant in the following study. In this case, the expressions for $\Delta_{2,\alpha\beta}^{*(2)}$ and $\Delta_{3.1,\alpha}^{*(2)}$ in 2D read

$$\Delta_{2,\alpha\beta}^{*(2)} = \frac{2\tau^2}{(1-b)^2}\left[(1-b)\rho\frac{\partial T}{\partial r_{<\alpha}}\frac{\partial T}{\partial r_{\beta>}} - \rho T b \frac{\partial^2 T}{\partial r_{<\alpha}\partial r_{\beta>}} - \rho T \frac{\partial u_\gamma}{\partial r_{<\alpha}}\frac{\partial u_\gamma}{\partial r_{\beta>}} \right.$$
$$\left. + \frac{T^2}{\rho}\frac{\partial \rho}{\partial r_{<\alpha}}\frac{\partial \rho}{\partial r_{\beta>}} - bT\frac{\partial T}{\partial r_{<\alpha}}\frac{\partial \rho}{\partial r_{\beta>}} - T^2 \frac{\partial^2 \rho}{\partial r_{<\alpha}\partial r_{\beta>}}\right], \quad (48)$$

$$\Delta_{3.1,\alpha}^{*(2)} = \frac{\tau^2}{1-b}p\left[(2+b)\frac{\partial T}{\partial r_\beta}\frac{\partial u_\beta}{\partial r_\alpha} + (6-3b)\frac{\partial u_\alpha}{\partial r_\beta}\frac{\partial T}{\partial r_\beta} - (6-3b)\frac{\partial u_\beta}{\partial r_\beta}\frac{\partial T}{\partial r_\alpha} \right.$$
$$\left. -2(1-b)T\frac{\partial^2 u_\beta}{\partial r_\alpha \partial r_\beta} + T\frac{\partial^2 u_\alpha}{\partial r_\beta \partial r_\beta}\right], \quad (49)$$

where the angle brackets in the subscript denote trace-free tensor, as an example $\frac{\partial^2 T}{\partial r_{<\alpha}\partial r_{\beta>}} = \frac{\partial^2 T}{\partial r_\alpha \partial r_\beta} - \frac{1}{2}\left(\frac{\partial^2 T}{\partial x^2} + \frac{\partial^2 T}{\partial y^2}\right)\delta_{\alpha\beta}$ for 2D. More details about trace-free tensor refer to the Appendix A 2.2 in literature[39].

## 5. Numerical results

In this section, we validate the proposed discrete ES-BGK model by several tests, including viscous sod shock tube, no slip thermal Couette flow, a Mach 3 wind tunnel with a step, and the steady shock wave. For the first two cases, the D2V19 model is adopted. The results can be used to verify the ability of the new model to describe different Prandtl number situations. For the latter two tests, the D2V36 model is adopted. The simulation of the Mach 3 wind tunnel with a step can be used to verify the applicability of the new model to the 2D shock wave cases and the calculation of steady shock wave shows the relations between viscous stress (heat flux) in Burnett level and the NOMF (NOEF).

## 5.1 Viscous sod shock tube

The initial conditions of sod tube are adopted, which reads

$$\begin{cases}(\rho,u_x,u_y,T)_L = (1,0,0,1),\\ (\rho,u_x,u_y,T)_R = (0.125,0,0,0.8).\end{cases} \quad (50)$$

The simulation area is divided into two parts, the subscript "$L$" and "$R$" in Eq.(50) are the initial condition of the left half part and right half part, respectively. The D2V19 model in Fig.1 (a) is adopted and $c = 2.0$. Simulations are carried out under the condition: the space step $\Delta x = \Delta y = 1\times 10^{-3}$, the number of mesh $Nx \times Ny = 1000\times 1$, the time step $\Delta t = 2\times 10^{-6}$, and the relaxation time $\tau = 2\times 10^{-4}$. The free inflow and free outflow boundary conditions are adopted in left and right, respectively. For all simulations in this paper, the second order NND scheme is adopted to solve the space derivation and the first order forward difference is used to solve the time derivation in Eq.(22).

Five cases with different kinds of Prandtl numbers are calculated. Figure 2 shows the results of velocity and temperature profiles at $t = 0.18$. The Riemann solutions (the solid line) are also plotted for comparison. The Riemann solutions are calculated based on Euler equations, which does not include the effects of viscosity and heat flux, while the simulation results (the symbols) contain those. So there exist distinct transition zones around the discontinuities for both velocity and temperature profiles of simulation results.

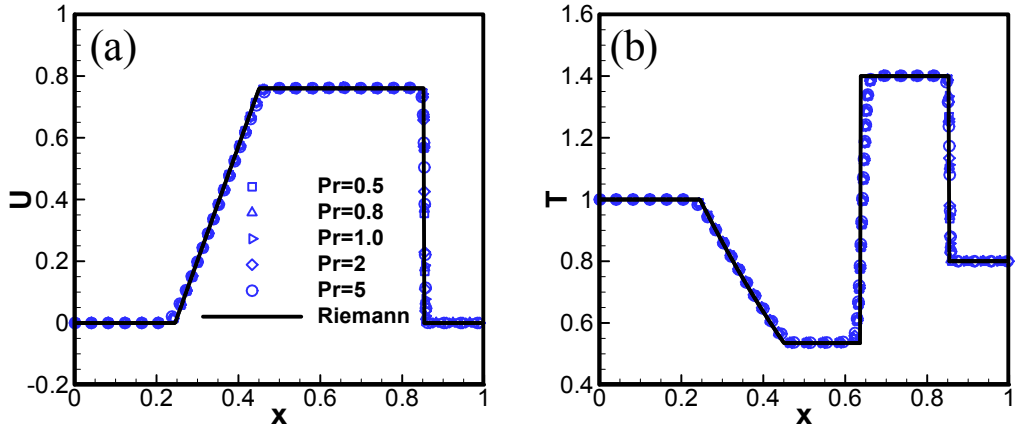

**Fig.2 Profiles of velocity and temperature of sod shock tube at $t = 0.18$: (a) velocity, (b)temperature. The simulation results are denoted by symbols and the Riemann solution is represented by solid lines.**

Figure 3 shows the profiles of viscous stress and heat flux. The NOMF and the NOEF are also plotted for comparison. The viscous stress and heat flux are calculated by Eqs.(40) and (41), respectively. The values of viscous stress show significant difference while the values of heat flux are the same for different Prandtl numbers. It can be observed that the values of NOMF are well agree with viscous stresses in NS level and the effect of Prandtl number is naturally included in NOMF.

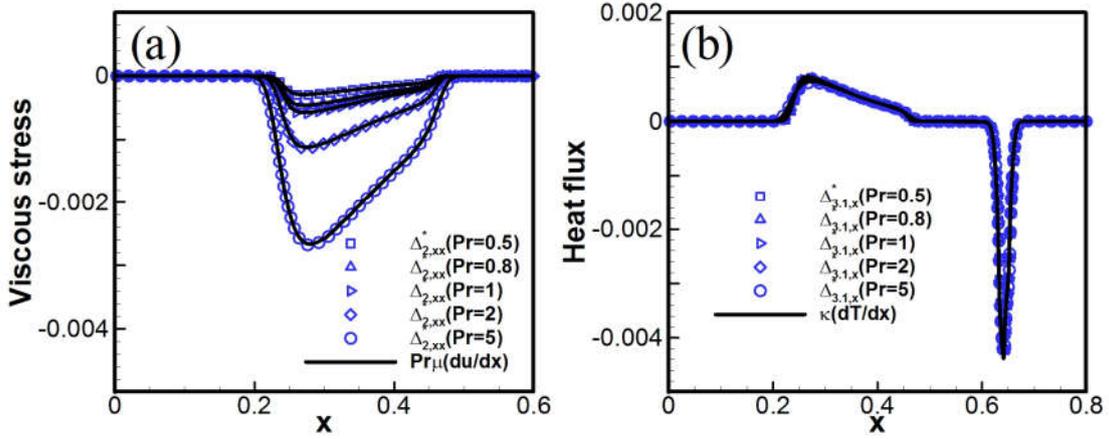

Fig.3 Profiles of viscous stress and heat flux of sod shock tube: (a) viscous stress, (b) heat flux. The NOMF and NOEF are denoted by symbols and the viscous stress and flux are represented by solid lines.

## 5.2 No slip thermal Couette flow

Couette flow with a temperature gradient is a good test to describe the viscous heat conducting flow. Consider a gas flow between two parallel walls with the bottom fixed and the top wall moving with a speed $U$ in the $x$ direction. The temperature of the bottom and the top wall are fixed with $T_0$ and $T_1$, respectively. According to incompressible NS equations the steady state analytic dimensionless temperature distribution reads[40, 41]

$$\frac{T-T_0}{T_1-T_0} = \frac{y}{H} + \frac{\Pr \mathrm{Ec}}{2}\frac{y}{H}(1-\frac{y}{H}), \tag{51}$$

where Ec is the Ecker number, $\mathrm{Ec} = \dfrac{U^2}{C_p(T_1-T_0)}$.

The D2V19 model in Fig.1(a) is adopted and $c=1.6$. The simulation distance of two walls is $H=1.0$ and the number of mesh is $Nx \times Ny = 1 \times 500$. The bottom wall is fixed with a temperature $T_0 =1.0$, while the moving velocity of the top wall is $U=0.2$ with a temperature $T_1 =1.001$ (i.e., Ec=20). The initial density of the gas inside the channel is 1.0 and the initial velocity is $u(y) = yU/H$, so the fluid in the channel can be approximated as incompressible. No slip wall boundary conditions are adopted to the bottom and the top[42]. Different Prandtl numbers and Ecker numbers are simulated and compared with analytical solutions. The Ecker number is adjusted by $T_1$, for example when $T_1 = 1.005$, Ec = 4.

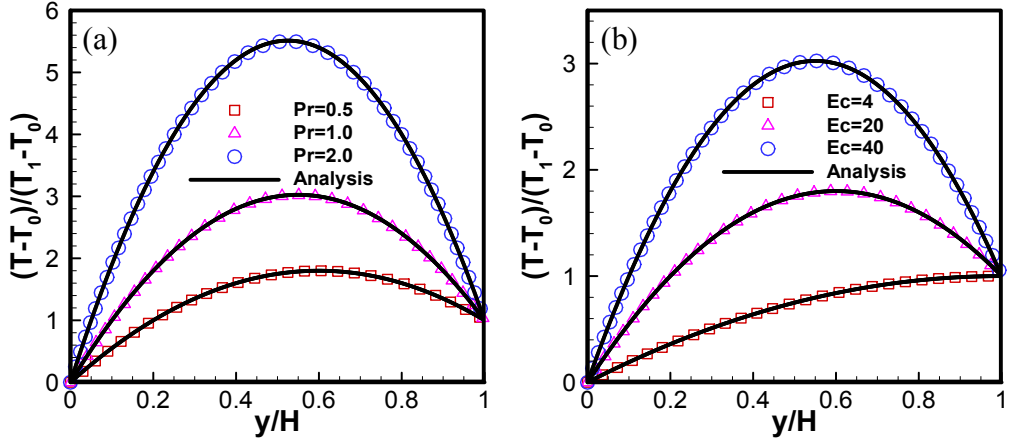

**Fig.4 The temperature spatial distribution on steady state for difference Prandtl numbers and Ecker numbers: (a) different Prandtl numbers, (b) different Ecker numbers. The solid lines denote the analytic solutions and the symbols are simulation results.**

Figure 4 gives the simulation results. Figures 4(a) and 4(b) present the results in the cases with different Prandtl numbers and Ecker numbers, respectively. The simulation results fit the analytic solutions very well, from which we can conclude that the Prandtl number correction in discrete ES-BGK does works well.

## 5.3 A Mach 3 wind tunnel with a step

This two-dimensional test is first introduced by Emery and is widely used to verify the capacity of the new model to capture shock wave in 2D case[43]. The problem begins with uniform Mach 3 flow in a wind tunnel containing a step. The tunnel has a length of 3 and a width of 1. The step is located at 0.6 from the entrance with a width of 0.2. Initially, the wind tunnel is filled with a gas with density $\rho_0 = 2.0$, pressure $p_0 = 1.0$, and velocity $u_0 = 3.0$. The inflow and the outflow boundary conditions are adopted in the left and the right boundary. Along the walls of the tunnel, reflecting boundary conditions are applied. As demonstrated in Ref.[43], the corner of the step is the center of a rarefaction and hence is a singular point of the flow. Just as a qualitative study, we have not done anything special at the singular point.

The D2V36 model in Fig.1(b) is used (with $c$=2.0) to ensure the symmetry of the discrete velocity which has significant effect on numerical stability of simulation, especially for 2D problem. The results are shown in Figs. 5. The contours of density and pressure are presented in Figs. 5(a) and 5(b), respectively. The results calculated by numerical solving NS equations are also plotted at the bottom half of the figures for comparison. It can be seen that the results of the new model are well consistent with those from NS equations. This simulation results show that the discrete ES-BGK with D2V36 model can be well used to simulate the two-dimensional high speed compressible problems.

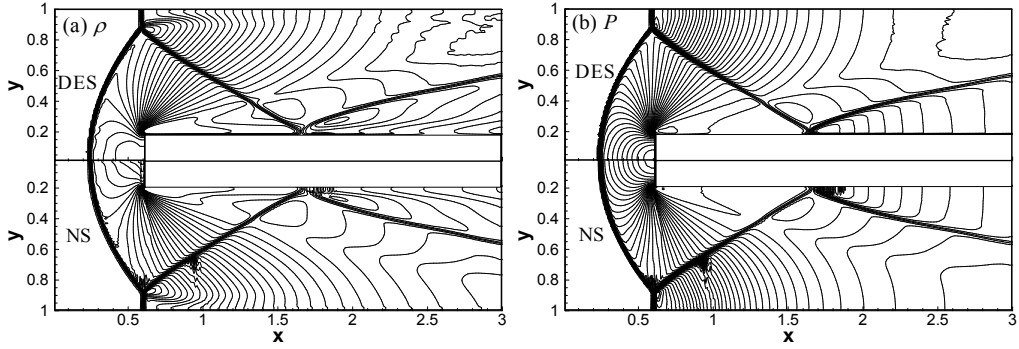

**Fig.5 The contours of density and pressure of steady state for Mach 3 step problem: (a)density contour, (b) pressure contour. The upper half of the figures are obtained by the discrete ES-BGK and the bottom half of the figures are results of NS.**

## 5.4 Non-equilibrium effects of the shock wave

In the above cases, the validation of the discrete ES-BGK model in NS level is verified. In fact, as we have mentioned, the new model also works in Burnett level. To verify the characteristic of nonlinear constitutive relation of the new model, one-dimensional steady shock wave is simulated and the viscous stress and heat flux around the wave front are presented and discussed. The initial state is set by Hugoniot relation with 1.5 Mach,

$$\begin{cases}(\rho,u_x,u_y,T)_L = (1.5882, 0.7857, 0, 1.6790),\\ (\rho,u_x,u_y,T)_R = (1.0, 0, 0, 1.0).\end{cases} \tag{52}$$

At beginning, the shock is located at $x = 0.02$. The D2V36 model in Fig.1(b) is adopted and $c=1.5$. The computation is carried out with the mesh number $Nx \times Ny = 5000 \times 1$. The space step is $\Delta x = \Delta y = 2 \times 10^{-4}$, time step is $\Delta t = 1 \times 10^{-6}$, and the relaxation time is $\tau = 1 \times 10^{-3}$. The free inflow and outflow boundary condition are adopted in left and right boundary, respectively.

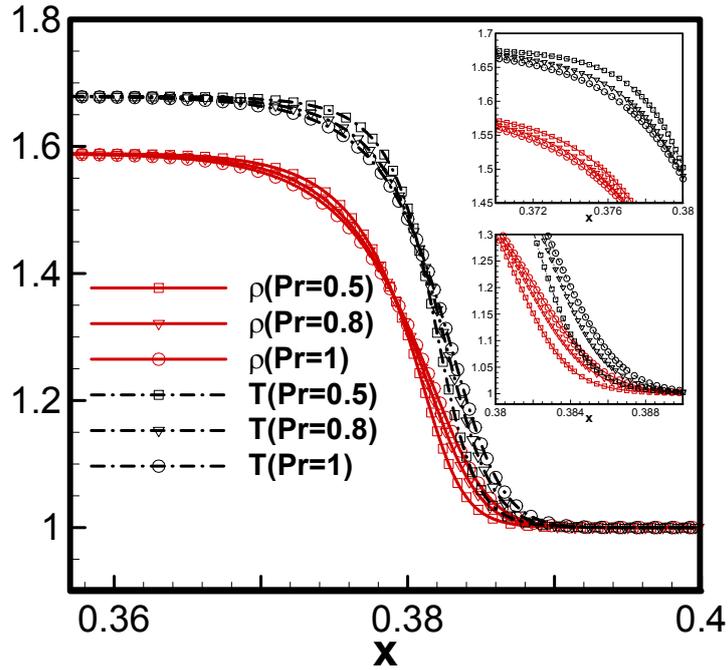

**Fig.6 Profiles of density and temperature in the steady state of a 1.5 Mach shock wave for difference Prandtl numbers. The solid lines with symbols are density profiles and the dash dot lines with symbols denote temperature profiles. Different Prandtl numbers are distinguished by different shape of symbols.**

Three different Prandtl numbers are simulated and the profiles of density and temperature in the steady state is presented in Fig.6. It can be found that profiles for different Prandtl numbers nearly intersect at one point around $x$=0.38 for both density and temperature. In front of the intersection ($x$< 0.38), the larger values of density and temperature correspond to the smaller Prandtl number, while, behind the intersection, the larger values of density and temperature correspond to the larger Prandtl number. The difference between the three Prandtl numbers is not significant in Fig.6, so the corresponding values of viscous stress and heat flux around shock front are given in Fig.7.

The profiles of the NS viscous stress (heat flux), the Burnett viscous stress (heat flux), and the NOMF (NOEF) are plotted in the same figure for comparison. From Fig.7 we can observe that the NOMF is well coincide with the NS and the Burnett viscous stress when the system is in near the equilibrium state. However, with the increase of the non-equilibrium effects, the NS viscous stress firstly deviate from NOMF while the Burnett viscous stress is still well coincide with NOMF. With the further increase of the non-equilibrium effect, the Burnett viscous stress is also deviate from NOMF around the maximum point. The characteristic of heat flux and NOEF is similar.

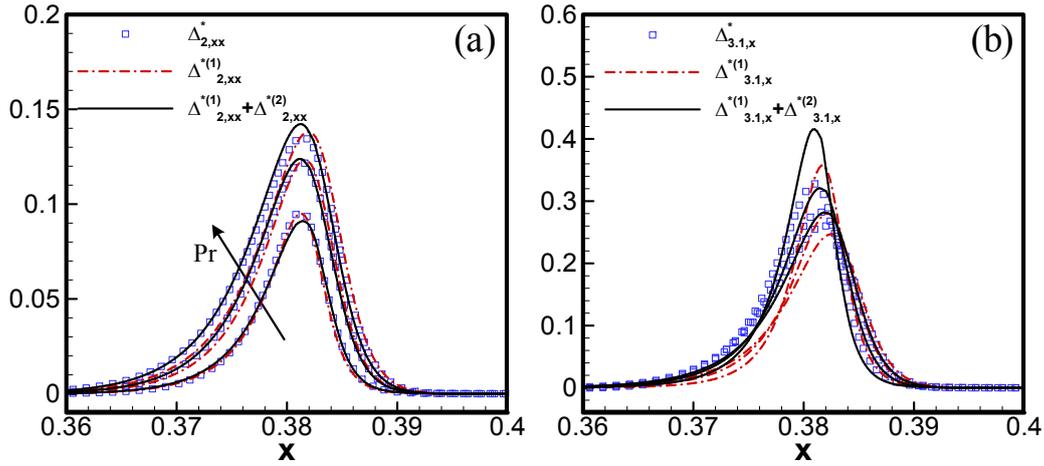

**Fig.7** Comparisons of viscous stress (heat flux) and NOMF (NOEF): (a) profiles of viscous stress and NOMF, (b) profiles of heat flux and NOEF. Dash dot lines are the NS viscous stress and heat flux, solid lines denote the Burnett viscous stress and heat flux, and the symbols represent the NOMF and NOEF. The arrow points to the increasing direction of the Pr.

From figure 7 we can concluded that the new model is in Burnett level. Taking account of the second order term, the viscous stress and heat flux are more approximate to the NOEF and NOMF than the first order ones when the non-equilibrium strength is much higher within a certain extent.

## 6. Recovering of the velocity distribution function

In Ref.[27] the qualitative information on the actual distribution function was drawn using the components of the non-equilibrium quantities, $\mathbf{\Delta}_m^*$. Recently, it has been found that the actual distribution function, to some extent, can be recovered quantitatively, although the discrete distribution function is not a real one in the evolutionary equation of DBM. The recovering of the velocity distribution function is often needed since the distribution function can be used to characterize the non-equilibrium state of flow system.

According to the CE expansion, the actual velocity distribution function can be expanded as Taylor series near the equilibrium state which read (for BGK model)

$$f = f^{eq} + \varepsilon f^{(1)} + \varepsilon^2 f^{(2)} + \cdots \qquad (53)$$

where $\varepsilon$ is a small quantity which has a positive correlation with the dimensionless relaxation time (or Knudsen number). When $\varepsilon \to 0$ which means the system reaches its equilibrium in an infinitely small amount of time, it seems that the velocity distribution function is always in its equilibrium state. With the increase of $\varepsilon$, such as $\varepsilon = 0.01$, the term with first order of $\varepsilon$ may not be ignored but the term with second order of $\varepsilon^2$ is still so insignificant that can be ignored and the truncated terms would not make much difference. The larger the value of $\varepsilon$, the more terms in

Eq.(53) need to be kept, and the more deviation of the actual distribution function away from its equilibrium state. Our discussions below are limited to the situation near equilibrium state.

Firstly, in the NS level, from the CE expansion we know that $f = f^{eq} + \varepsilon f^{(1)}$ and

$$\varepsilon f^{(1)} = -\tau \left( \varepsilon \frac{\partial f^{eq}}{\partial t_1} + v_\alpha \varepsilon \frac{\partial f^{eq}}{\partial r_{1\alpha}} \right). \tag{54}$$

Since the equilibrium distribution function $f^{eq}$ can be expressed by $\rho, u_\alpha$, and $T$, the derivative of $f^{eq}$ can turn into the derivative of $\rho, u_\alpha$, and $T$.

$$f^{(1)} = f^{eq} \left[ D_\rho \left( \frac{\partial \rho}{\partial t_1} + v_\alpha \frac{\partial \rho}{\partial r_{1\alpha}} \right) + D_T \left( \frac{\partial T}{\partial t_1} + v_\alpha \frac{\partial T}{\partial r_{1\alpha}} \right) + D_{u_\beta} \left( \frac{\partial u_\beta}{\partial t_1} + v_\alpha \frac{\partial u_\beta}{\partial r_{1\alpha}} \right) \right] \tag{55}$$

with $D_\rho = \frac{1}{\rho}$, $D_T = \left[ -\frac{D}{2}\frac{1}{T} + \frac{|v_\alpha - u_\alpha|^2}{2RT^2} \right]$, and $D_{u_\beta} = -\frac{(v_\beta - u_\beta)}{RT}$. In addition, the time partial derivatives in Eq.(55) can be represented by the space partial derivative through Euler equations, which read

$$\frac{\partial \rho}{\partial t_1} = -\rho \frac{\partial u_\alpha}{\partial r_{1\alpha}} - u_\alpha \frac{\partial \rho}{\partial r_{1\alpha}}, \tag{56}$$

$$\frac{\partial u_\alpha}{\partial t_1} = -\frac{T}{\rho} \frac{\partial \rho}{\partial r_{1\alpha}} - \frac{\partial T}{\partial r_{1\alpha}} - u_\beta \frac{\partial u_\alpha}{\partial r_{1\beta}}, \tag{57}$$

$$\frac{\partial T}{\partial t_1} = -u_\alpha \frac{\partial T}{\partial r_{1\alpha}} - \frac{2}{D} T \frac{\partial u_\alpha}{\partial r_{1\alpha}}. \tag{58}$$

Then the first order approximation of the actual velocity distribution function $f$ is

$$f \approx f^{eq} + \varepsilon f^{(1)} = f^{eq} \left( 1 + D_\rho \left( \frac{\partial \rho}{\partial t} + v_\alpha \frac{\partial \rho}{\partial r_\alpha} \right) + D_T \left( \frac{\partial T}{\partial t} + v_\alpha \frac{\partial T}{\partial r_\alpha} \right) + D_{u_\beta} \left( \frac{\partial u_\beta}{\partial t} + v_\alpha \frac{\partial u_\beta}{\partial r_\alpha} \right) \right), \tag{59}$$

where $\frac{\partial}{\partial t} = \varepsilon \frac{\partial}{\partial t_1}$ and $\frac{\partial}{\partial r_\alpha} = \varepsilon \frac{\partial}{\partial r_{1\alpha}}$. For example, at the point $x = 0.38$ for Pr=1 in Fig.6, the $\rho, u_\alpha, T$, and their corresponding time partial derivatives and space partial derivatives are calculated and shown in Table 1.

As a result, the actual distribution function at point $x=0.38$ is obtained. The actual distribution function and its corresponding contour is shown in Fig.8. The corresponding equilibrium distribution function is also plotted on the right hand (i.e. figures 8(b) and 8(d)) for comparison. It can be found that the actual distribution function has two peak in the *x* direction which is a typical characteristics of shock waves[44].

**Table 1 The macroscopic quantities and their derivatives at point x=0.38 for Pr=1 in Fig.6.**

| $\rho$ | $u_x$ | $T$ | $\dfrac{\partial \rho}{\partial x}$ | $\dfrac{\partial u_x}{\partial x}$ | $\dfrac{\partial T}{\partial x}$ | $\dfrac{\partial \rho}{\partial t}$ | $\dfrac{\partial u_x}{\partial t}$ | $\dfrac{\partial T}{\partial t}$ |
|---|---|---|---|---|---|---|---|---|
| 1.3190 | 0.5130 | 1.4858 | -50.50 | -61.55 | -49.00 | 107.09 | 137.46 | 116.59 |

Similarly, the second order or higher order approximation of the actual distribution function can also be recovered in the same way if it is necessary. It should be noted that the second order space derivatives will be needed for second order approximation of the actual distribution function and the *m*-th order space derivatives will be needed for the *m*-th approximation. Of course, the accuracy of the corresponding space derivatives need to be guaranteed. The actual distribution of discrete ES-BGK model can also be recovered by taking account the $f^{ES}$ terms. Besides, this idea of recovering the actual distribution function by macroscopic quantities and their space derivatives is also applied for NS equations and Burnett equations.

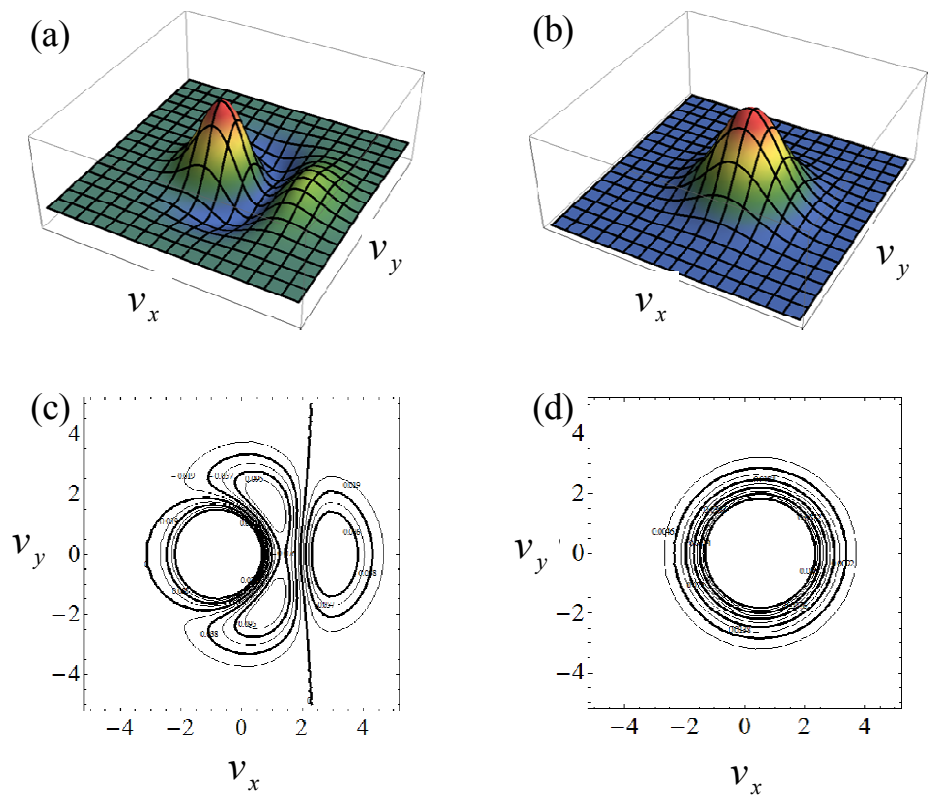

**Fig.8 The velocity distribution functions and their contour plot. (a) and (c) are first order approximation of the actual distribution function and its contour plot. (b) and (d) are the corresponding local equilibrium distribution function and contour plot.**

## 7.Conclusion

Under the framework of the CE multiscale analysis, a scheme for constructing discrete ES-BGK model is presented. Compared with the original discrete BGK model, the new model works in a wider range of Knudsen number and has a flexible Prandtl number. As an example, the discrete ES-BGK model in Burnett level is illustrated; the viscous stress and heat flux in Burnett equations are related to the non-equilibrium quantities. The new model is verified by four numerical tests. When the system is near its thermodynamic equilibrium state, the NS viscous stress and heat flux works well. With the increasing of non-equilibrium strength, the NS viscous stress and heat flux should be replaced by the Burnett ones. When the non-equilibrium strength further increase, the viscous stress and heat flux should include higher order non-equilibrium effects. Correspondingly, more kinetic moments of equilibrium distribution function are generally needed in the model construction. In addition, a new scheme is proposed to quantitatively recover the main feature of actual velocity distribution function from the macroscopic quantities and their space derivatives. Such a recovery method is needed when the distribution function is used to characterize the non-equilibrium state of flow system. It works not only for simulations based on DBM but also for those based on hydrodynamic models, such as the NS and Burnett equations.

## Acknowledgments:

The authors would like to acknowledge support of National Natural Science Foundation of China [under grant nos. 11475028,11772064, 11502117, and U1530261], Science Challenge Project (under Grant No. JCKY2016212A501 and TZ2016002).